# Thermal rectification due to phonon confinement in nanoparticles

Alexej D. Semenov[1], Mariia Sidorova[2], Alessio Zaccone[3], and Marcel Di Vece[3, 4]

[1] DLR Institute of Space Research, Berlin, Germany

[2] PTB Berlin, Germany

[3] Physics Department "Aldo Pontremoli", Università degli Studi di Milano, Via Celoria 16, 20133, Milan, Italy

[4] Interdisciplinary Centre for Nanostructured Materials and Interfaces (CIMaINa)

## ABSTRACT

We demonstrate that thermal rectification can arise at the contact between two spherical nanoparticles of identical material but different size due to the geometric confinement of phonons. This confinement suppresses long-wavelength phonons differently in differently sized particles and creates a size-dependent gap in the phonon density of states. This gives rise to direction-dependent heat transport even in perfectly homogeneous materials. We develop an analytical model based on phonon confinement and phonon ray-tracing in the Casimir regime and derive expressions for heat fluxes and rectification efficiency as functions of particle sizes and temperatures. The model predicts measurable rectification efficiencies for nanoparticles with radii of a few tens of nanometers, reaching fraction of percent at room temperature and much larger values at low temperatures. The proposed mechanism provides a straightforward and scalable route to thermal rectification in granular nanomaterials without requiring material heterogeneity or strong nonlinearities.

## 1. INTRODUCTION

Thermal rectification (asymmetric heat transport under reversed temperature gradients) is essential for thermal diodes, refrigeration, and heat management in nanoscale devices. Achieving rectification without material (e.g. chemical) heterogeneity or complex nonlinear structures remains a major challenge, particularly in nanoparticle-based systems.

In bulk systems, thermal rectification has been demonstrated by combining materials with strongly non-linear and different temperature dependences of the thermal conductivity [1,2]. At reduced length scales, new mechanisms become available that exploit nonlinear and asymmetric structures, including geometric asymmetry, asymmetric doping and functionalization, see review [3]. These strategies have enabled thermal switches [4] and thermal transistors [5]. Another rectification mechanism occurs in electronic systems, where heat is carried by electronic quasiparticles rather than phonons. [6]. Rectification occurs as the direction-dependent flow of hot quasiparticles across a superconductor-normal-metal interface due to the energy gap in the superconducting density of states. This mechanism was successfully used to cool metallic nanobolometers below the ambient phonon temperature [7].

The theoretical description of phononic rectifiers depends on the heat transport regime. On macroscopic scales, larger than the bulk mean free path of phonons, $l_{MPF}$, the thermal conductivity $(\kappa)$ is described by the kinetic theory as $\kappa = \frac{1}{3} c_{\mathrm{D}} l_{\mathrm{MPF}} u$, relating it to the Debye phonon heat capacity $c_{\mathrm{D}}$, and sound velocity $u$. In this regime, heat flow through an interface is modelled using Fourier's law with appropriate boundary conditions [8,9]. However, at the micro/nano scale, the device dimensions can become comparable or smaller than the phonon mean free path. In this case, phonon scattering is dominated by boundaries, placing the system in the so-called Casimir regime where phonons propagate ballistically between boundaries. In this regime, an effective mean free path can be introduced, allowing Fourier's law to remain formally applicable (see e.g. [10,11]).

In free nanoparticles (NPs), vibrational modes are considerably modified and their total number is drastically reduced [12]. Strictly speaking the remaining modes differ from plane propagating acoustic waves, but one can still apply the concept of the phonon mode confinement to the Debye phase space which was done for two-dimensional nanofilms to obtain the heat capacity [13] and the rate of heat removal [14]. For one-dimensional nanowires, the heat flow and rectification efficiency are commonly modelled with the molecular dynamic simulation technique [16].

In this work, we show that thermal rectification can arise solely from size-dependent geometrical phonon confinement in NPs of identical materials. We developed an analytical approach for estimating the rectification strength at the contact between two nanospheres of different diameters. By combining phonon confinement with heat transport in the Casimir regime and phonon ray tracing [17], we demonstrate that direction-dependent heat flow emerges from the confinement-induced gap in the phonon density of states.

Importantly, the estimated rectification efficiency reaches experimentally measurable values already at room temperatures for spherical nanoparticles with diameters of a few tens of nanometers, in agreement with recent experiments on aggregates of Ge nanoparticles [18]. The gap in the confined phonon density [13,14,19] motivates the discussion of possible phonon-based refrigeration in granular two-dimensional films. Ultimately, the physical origin of rectification is the confinement-induced modification of the phonon density of states, which creates direction-dependent spectral filtering of phonon flux.

## 2. Physical model and assumptions

Consider two spherical NPs of identical crystalline material but different radii in thermal contact. Heat transport inside each particle is assumed to occur in the Casimir regime, in that the bulk phonon mean free path, $l_{MPF}$, exceeds the sphere diameter. In this regime, phonon scattering at the NP boundaries dominates, leading to rapid thermalization, so that phonons in each particle are characterized by the Bose distribution with a well-defined temperature affiliated with the temperature of the NP surface. Since the material is identical, differences in forward and reverse heat transport arise solely from geometrical confinement rather than intrinsic material properties.

The thermal contact between nanospheres is assumed weak and non-epitaxial, e.g., mediated by van der Waals forces and/or covalent bonds, with disrupted crystalline order at the interface that allows for temperature discontinuity. Phonons incident on the contact from both sides are absorbed and re-emitted into both NPs without preserving spectral identity. Heat exchange between NPs occurs through an effective contact area $S,$ determined by the microscopic separation between the spheres and by the ratio of the thermal conductivities of the embedding medium and nanospheres (see Ref. [9] for a detailed discussion). In experimentally relevant situations, such as Ge nanoparticles embedded in dry air [18], this area satisfies $S \ll R_1^2; R_2^2$ where $R_1$ and $R_2$ are NPs radii. All microscopic interface details are absorbed into a dimensionless, angle-independent energy transparency $\alpha$, which for simplicity is assumed identical for both directions and independent of temperature and sphere size.

We adopt a black-body–like analogy in which the contact area $S$ emits towards both NPs phonons with Debye spectrum at an effective temperature $T_{\mathrm{x}}$. However, the flux entering the NP is limited by the phonon spectrum of this particle. Similarly, heat flux hitting the contact area from the NPs side is also limited by the phonon spectrum of this NP. Temperature of the contact area, $T_{\mathrm{x}}$, stays between temperatures of contacting NPs and plays the role of a self-adjusting parameter ensuring continuity of heat flow.

For a dimer of two different spherical NPs (Fig. 1), heat exchange is controlled by the difference between opposing heat fluxes from NP2 to NP1 and from NP1 to NP2. The asymmetry in sphere sizes generates rectification, despite identical material composition, because different radii impose different phonon confinement. Since phonon confinement is stronger in the smaller particle, for the same temperature gradient, the forward heat flow from the smaller to the larger particle is larger than the reversed heat flow in the opposite direction.

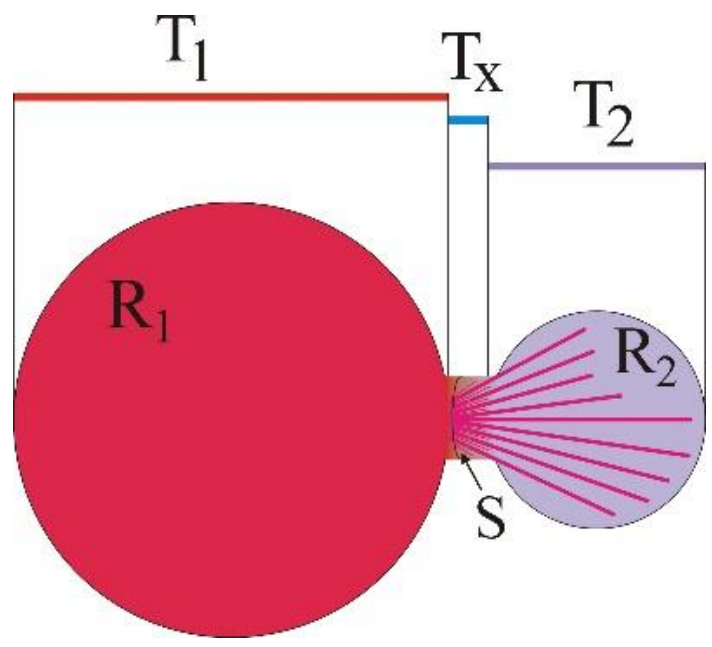

**Fig. 1**. *Two nanospheres of different radii, $R_1$ and $R_2$, at different temperatures, $T_1$ and $T_2$, in thermal contact via the area $S$ (not in scale) with the temperature $T_x$. Straight red lines depict phonon flux from the contact area into the smaller right sphere.*

Phonon confinement occurs when the sphere radius is comparable to the mean phonon wavelength $\approx \frac{0.37hu}{k_B T}$ (3D Debye model, where $u$ is the phonon velocity, $T$ the temperature, $k_B$ and $h$ are Boltzmann and Plank constants, respectively). This suppresses long-wavelength modes and discretize the otherwise continuous 3D Debye density of phonon states [13,14,19]. For a sphere of radius $R$ much larger than the crystalline unit cell $a_0$, surface vibrational modes can be neglected and a semiclassical ray-tracing description of phonon transport is justified [17]. On the other hand, for $a_0 \ll R$, one can neglect discretization of $k$ -values and use the continuous Debye formalism.

The visualization of phonon-confinement is clear-cut in reciprocal, wavevector space. For an isolated spherical nanoparticle of radius $R$, confinement is isotropic and suppresses phonon modes with wavelengths $\lambda > 2R$, corresponding to a lower cutoff wavevector $k_{\min} = \frac{\pi}{R}$. In reciprocal $k$-space, this excludes from the Debye sphere a spherical region with $k < k_{\min}$ around the origin.

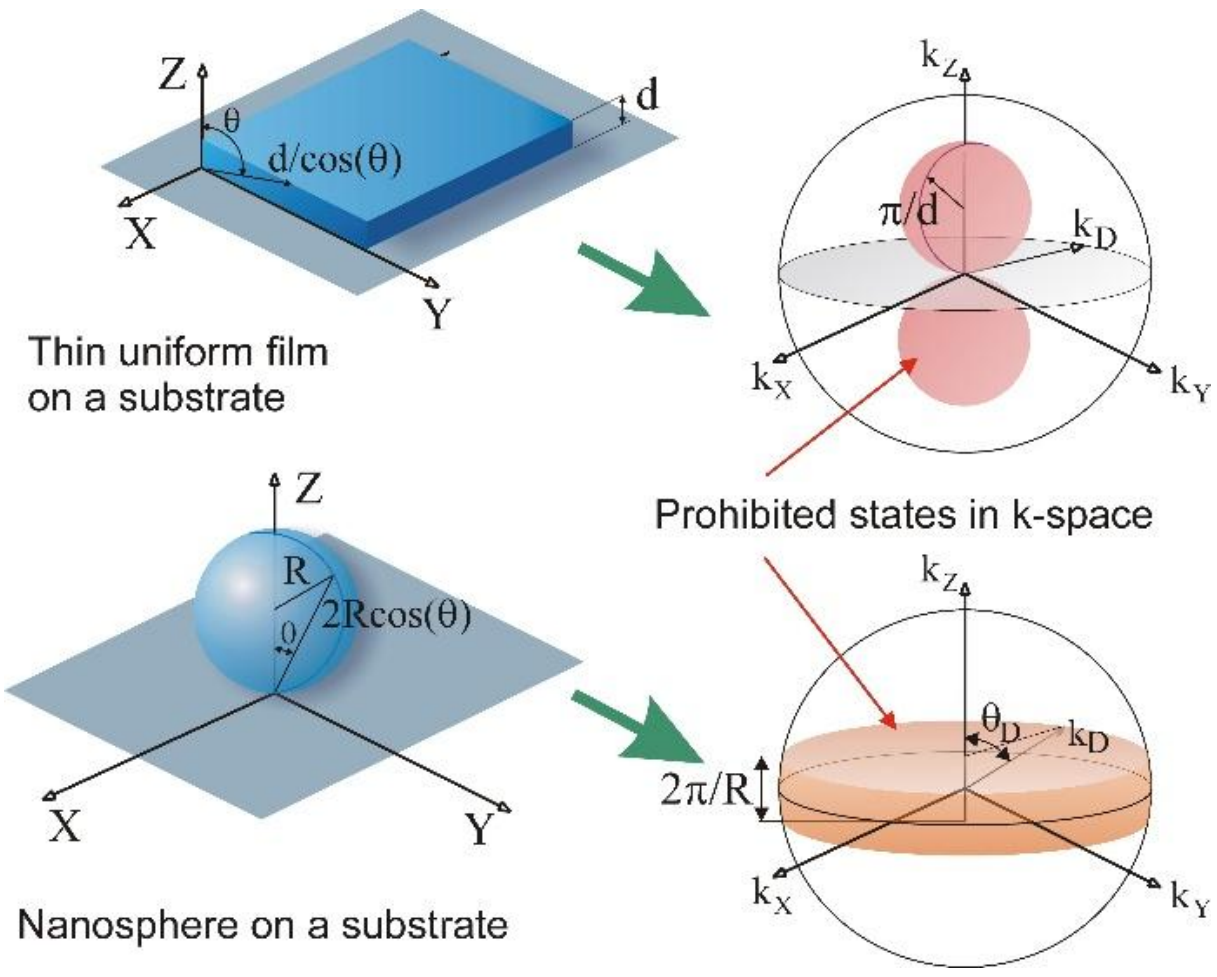

**Fig.2.** *Confinement in real and reciprocal space. Left side: Real-space schematics of a thin film of thickness $d$ on an infinite substrate (top) and a nanosphere with the radius R in thermal contact with the substrate (bottom). Right side: In reciprocal, wavevector k-space, which is a sphere of Debye radius $k_D$, confinement in the film introduces two empty spherical cavities with radii $\frac{\pi}{d}$ while confinement in the nanosphere introduces an empty spherical slab of thickness $\frac{2\pi}{R}$.*

The effect of confinement on the heat exchange depends on the geometry and location of the thermal contact. For comparison (illustrative only), confinement in a thin film of thickness $d$ on a substrate equally affects all phonons in the film propagating at the same angle [14]. Phonon modes with angle-dependent wavelengths $\lambda(\theta) > d\ \cos\theta$ are suppressed, resulting in a cutoff $k_{\min}(\theta) = \frac{2\pi\cos\theta}{d}$. In *k*-space, this excludes two spherical regions tangent at the origin of the Debye sphere [13,14] (Fig. 2, top).

For an NP with the radius $R$ on the substrate (Fig. 2, bottom), heat exchange occurs through a small almost $R$-independent contact area. Heat exchange in such geometry is supported only by phonons hitting this area and is proportional to the area itself. In turn, phonons hitting the contact from the NP side, only populate modes with $\lambda(\theta) < 2R\cos\theta$ (polar angle $\theta$ measured from the sphere diameter crossing the contact), corresponding to a cut-off $k_{\min}(\theta) = \frac{\pi}{R\cos\theta}$. In the Debye *k*-space, this excludes from heat exchange a spherical slab (confinement-induced gap) of thickness $\frac{2\pi}{R}$ symmetric with respect to the origin of the Debye sphere.

## 3. Thermal rectification between unequal nanospheres

In this work, we focus on a dimer formed by two spherical NPs of identical material but unequal radii, $R_1$ and $R_2$, in thermal contact as shown in Figs. 1 and 3. If NPs are kept at temperatures $T_1$ and $T_2$, respectively, net heat exchange is controlled by the difference between opposing phonon fluxes arriving from the contact area with the temperature $T_x$ into neighbors and fluxes arriving from NPs to the contact area. Fluxes in opposite direction are constrained by different receiving spheres, creating asymmetric available *k*-space for forward (from $R_2$ to $R_1$) versus reverse (from $R_1$ to $R_2$) temperature gradients. This asymmetry generates thermal rectification.

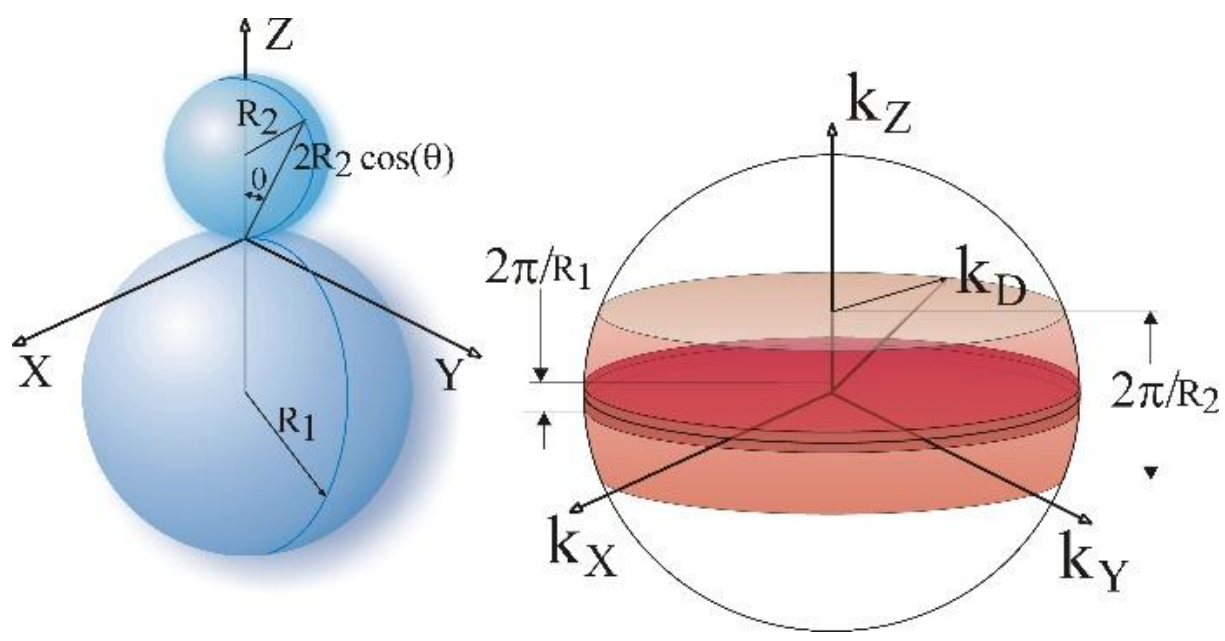

**Fig. 3**. *A dimer of two nanospheres of different radii, $R_1$ and $R_2$, in thermal contact (left) and reciprocal k-space (right). Phonon spectra allowed in the two nanoparticles differ due to size-dependent phonon confinement, resulting in direction-dependent heat transport.*

To calculate the energy flux from NP1 into NP2, we first integrate the Debye phonon flux received by the contact area from the sphere with radius $R$ and temperature $T$ over the confinement-modified *k*-space of the sphere. In thermal equilibrium, $T_x = T$, and the sphere receives the same flux from the contact area. Hence, the net heat flow becomes zero. The standard Debye flux (density of states $\propto k^2/u$, energy $\propto \hbar uk$, the Bose occupation $[e^{\frac{\hbar uk}{k_B T}} - 1]^{-1}$) is restricted by the angle-dependent cutoff $k_{min}\ (\theta) = \frac{\pi}{R\cos\theta}$ derived in Section 2. We introduce dimensionless variables: $x = k\frac{\hbar u}{k_B T}$, $x_D = k_D\frac{\hbar u}{k_B T}$, where $k_D = \left(\frac{6\pi^2}{a_0^3}\right)^{\frac{1}{3}}$ is the Debye wavevector, and $x_0(\theta, R) = \frac{\pi}{R\cos\theta}\frac{\hbar u}{k_B T}$, and denote by $\theta_D(R) = \arccos\left(\sqrt[3]{\frac{\pi}{6}}\frac{a_0}{R}\right)$ the angle at which the confinement-induced gap of thickness $\frac{2\pi}{R}$ intersects the surface of the Debye sphere.

The heat flux (power per unit contact area and per polarization in units $Wm^{-2}$) becomes [14, 20]:

$$P(T,R) = \frac{\alpha\,\hbar}{4\pi^2 u^2}\left(\frac{k_B T}{\hbar}\right)^4 \int_0^{\theta_D(R)} \sin\theta\cos\theta \int_{x_0(\theta,R)}^{x_D} \frac{x^3}{e^x - 1}\,\mathrm{d}x\,\mathrm{d}\theta. \quad (1)$$

We assume isotropic phonon dispersion with an identical sound velocity for all polarizations within the Debye approximation. However, because of the discontinuity of the shear modulus at the interface, it is more likely that only longitudinal phonons are involved in the heat transport between NPs. Due to confinement, the flux $P(T,R)$ explicitly depends on the radius, $R$, of the NP through the confinement-induced integration limits $0 < \theta < \theta_D(R)$ and $x_0(\theta,R) < x < x_\mathrm{D}$.

Let the two nanospheres have radii $R_1 > R_2$ and temperatures $T_1 = (T + \delta T)$ and $T_2 = T$, respectively, where $\delta T$ is the imposed temperature difference. The net heat flow in the reverse direction $q_{1\to 2}$, i.e. the heat flow through the contact area from the hotter, larger sphere to the colder, smaller one, is $q_{1\to 2} = P(T+\delta T, R_1) - P(T_\mathrm{x}, R_1)$. Where $P(T_\mathrm{x}, R_1)$ is the energy flux from the contact area into the larger sphere. Due to continuity reason, $q_{1\to 2}$ equals the net heat flow from the contact area to the smaller sphere $q_{1\to 2} = P(T_\mathrm{x}, R_2) - P(T, R_2)$. Although heat transport occurs in the ballistic Casimir regime, for $\delta T \ll T$, an effective thermal conductivity can be introduced adopting the center-to-center distance $R_1 + R_2$ as an effective mean free path [10]. Using Fourier's law for the heat flux (in units $Wm^{-2}$) $q_{1\to 2} = -\kappa_{1\to 2}(T)\frac{dT}{dx}$ with $\frac{dT}{dx} \simeq \frac{\delta T}{R_1+R_2}$, we define an effective reverse thermal conductivity (in units $WK^{-1}m^{-1}$) as

$$\kappa_{1\to 2} = \frac{P(T+\delta T, R_1) - P(T_\mathrm{x}, R_1)}{\delta T}(R_1 + R_2) = \frac{P(T_\mathrm{x}, R_2) - P(T, R_2)}{\delta T}(R_1 + R_2). \quad (2)$$

Setting the two sides of Eq.(2) equal to each other yields the condition for finding $T_\mathrm{x}$. For the forward direction, $\kappa_{2\to 1}$, reversing the temperature gradient exchanges the radii $R_1$ and $R_2$. Correspondingly, temperatures $T_\mathrm{x}$ for the forward and for the reverse directions differ. The equation embeds rectification through the radius-dependent allowed *k*-space. For $R_1 = R_2$ and $\delta T \ll T$, rectification disappears and $T_\mathrm{x} = T + \delta T/2$.

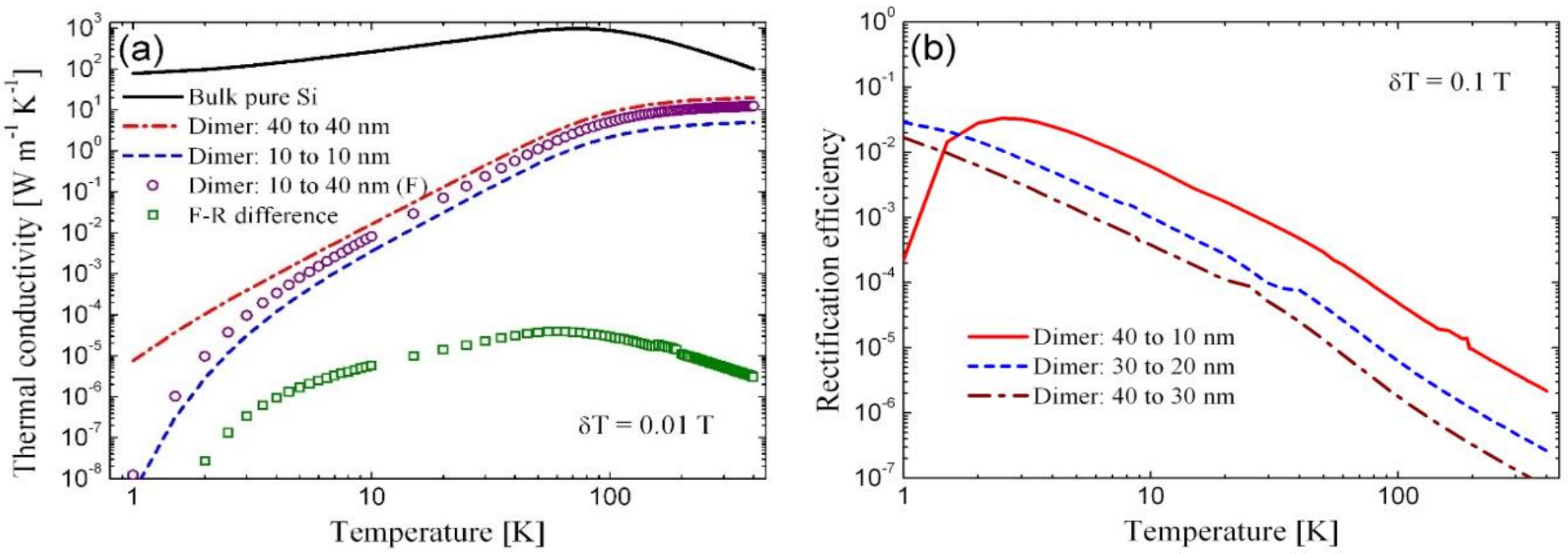

**Fig. 4.** *Thermal conductivities (a) and rectification efficiencies (b) under phonon confinement in dimers from Ge nanospheres. (a) Forward thermal conductivity* $\kappa_{2\to 1}$ *(open circles) and the difference* $\kappa_{2\to 1} - \kappa_{1\to 2}$ *(open squares) computed with Eq. (2) for a dimer with radii* $R_1 = 40$ *and* $R_2 = 10$ *nm,* $\delta T = 0.01\,T$ *and* $\alpha = 1$. *For comparison, shown are conductivities of two symmetric dimers* ($R_1 = R_2$) *with radii 40 (dash-dotted line) and 10 (dashed line) nm and experimental thermal conductivity of pure Ge [21] (solid line). (b) Rectification efficiency computed with Eq. (2.1) for dimers with radii (nm): 40 and 10; 30 and 20; 40 and 30 (Decoding in the legend) and* $\delta T = 0.1\,T$.

Figure 4 (a) illustrates forward, $\kappa_{2\to 1}$ conductivity for a dimer from Ge nanospheres (material parameters in Appendix) with radii $R_1$=40 nm, $R_2$=10 nm. Also shown is the conductivity difference $\kappa_{2\to 1} - \kappa_{1\to 2}$ between forward and reverse directions. For comparison, the plot shows heat conductivities for dimers with equal radii of 10 and 40 nm. In order to have thermal conductivities independent on $\delta T$, at each temperature, we choose $\delta T = 0.01\,T$. At high temperatures confinement is less pronounced because the phonon spectrum is shifted to higher energies. Here the forward conductivity of an asymmetric dimer approaches the conductivity of the larger, symmetric dimer. Contrarily, at low temperatures it is close to the conductivity of the smaller sphere with a stronger confinement. Overall, the thermal conductivity in dimers is much smaller than the conductivity of bulk Si and demonstrates a much stronger temperature dependence. Different, nonlinear temperature dependences of conductivities in symmetric dimers from different NPs provide the basis [2] for rectification in chains of nanospheres.

To describe the effect quantitatively, we introduce the rectification efficiency

$$\mathrm{Rec}(T) = \frac{\kappa_{2\to1}(T)-\kappa_{1\to2}(T)}{\kappa_{2\to1}(T)} = \frac{q_{2\to1}(T)-q_{1\to2}(T)}{q_{2\to1}(T)}, \qquad (2.1)$$

defined such that an ideal thermal diode with, e.g., $\kappa_{1\to2} = 0$ ($q_{1\to2} = 0$) will have 100% efficiency.

As shown in Fig. 4b, $\mathrm{Rec}(T)$ gradually decreases with temperature. At high contrast, $\frac{R_1}{R_2} = 40/10$, there appears a maximum at cryogenic temperatures. Reducing the contrast between the nanosphere sizes from $\frac{R_1}{R_2} = 40/10$ to 40/30 (upper and lower curves in Fig. 4b, respectively) strongly suppresses $\mathrm{Rec}(T)$ over the entire temperature range, illustrating that rectification is controlled primarily by different phonon confinement in different nanospheres.

While we illustrate these results using Ge parameters, the rectification mechanism stems from phonon confinement within the Debye model and is therefore universal to non-metallic NPs. Calculations for Si, $TiO_2$, and $HfO_2$ (Appendix A) yield quantitatively similar efficiencies, confirming this universality.

To isolate the rectification mechanisms, we expand heat fluxes $P(T,R)$ to the second order for small $\delta T \ll T$ to get $P(T+\delta T,R) = P(T,R) + P'(T,R)\delta T + 0.5P''(T,R)\delta T^2$ where prime refers to the temperature derivative. The reverse heat flow $q_{1\to2} = P(T+\delta T,R_1) - P(T_x,R_1) = P(T_x,R_2) - P(T,R_2)$ is then

$$q_{1\to2} = P'(R_1)\delta T + \frac{P''(R_1)}{2}\delta T^2 - P'(R_1)dT - \frac{P''(R_1)}{2}dT^2 = P'(R_2)dT + \frac{P''(R_2)}{2}dT^2. \qquad (2.2)$$

Here we omit the common temperature $T$. Solving the second part for $dT = T_x - T$ and substituting the solution back to Eq. (2.2), we obtain the reverse heat flow $q_{1\to2}$ and corresponding thermal conductivity according to Eq. (2) as $\kappa_{1\to2} = q_{1\to2}2\bar{R}\delta T^{-1}$ with $\bar{R} = (R_1+R_2)/2$. Forward flow $q_{2\to1}$ and thermal conductivity $\kappa_{2\to1}$ are obtained in the similar manner. Net heat fluxes for the reverse and forward directions are

$$q_{1\to2} = \frac{P'(R_1)P'(R_2)\delta T}{P'(R_1)+P'(R_2)} + \frac{1}{2}\frac{P''(R_2)P'(R_1)\delta T^2}{[P'(R_1)+P'(R_2)]^2}, \qquad q_{2\to1} = \frac{P'(R_1)P'(R_2)\delta T}{P'(R_1)+P'(R_2)} + \frac{1}{2}\frac{P''(R_1)P'(R_2)\delta T^2}{[P'(R_1)+P'(R_2)]^2}. \qquad (2.3)$$

The first terms in the expressions above are symmetric under gradient reversal and, for $R_1 = R_2$, represents ordinary heat transport between identical nanospheres $q_{1\to2} = q_{2\to1} = 0.5P'(R)\delta T$. The second term is asymmetric with respect to interchange of $R_1$ and $R_2$ and provides rectification. Note that this effect is non-linear (quadratic in $\delta T$). The asymmetry originates from confinement-controlled integration limits in Eq. (1).

Finally, substituting heat fluxes $q_{1\to2}$ and $q_{2\to1}$ from Eq. (2.3) in Eq. (2.1), we obtain rectification efficiency for $\delta T \ll T$

$$\mathrm{Rec}(T) = \frac{\delta T}{2}\frac{P'(R_1)P'(R_2)}{P'(R_1)+P'(R_2)}\left[\frac{P''(R_1)}{P'(R_1)^2} - \frac{P''(R_2)}{P'(R_2)^2}\right] \approx \frac{\delta T}{4}\frac{\frac{\partial}{\partial \bar{R}}[P''(T,\bar{R})]}{P'(T,\bar{R})}\delta R. \qquad (2.4)$$

The expression on the right-hand side is obtained by the first order expansion for $\delta R = R_1 - R_2 \ll R_1, R_2$. It closely resembles the result obtained in [2] if one realizes that $P''(R)$ is proportional to the heat capacity of a spherical nanoparticle with radius $R$.

## 1. Rectification in chains of unequal nanospheres

We now extend the dimer analysis to two linear chains, each composed of identical nanospheres, where the two chains differ only by their sphere radii and lengths, as shown in Fig. 5a. The physical mechanism of rectification in this geometry differs. For a dimer, rectification originates from the confinement-induced asymmetry in the phonon k-space of two different spheres. In contrast, when the chains are connected in series and their lengths are much larger than the sphere diameters, contribution of the single rectifying contact becomes progressively small with increasing chain lengths. Nevertheless, as we show below, rectification can persist in contacting long chains due to different non-linear temperature dependences of their effective thermal conductivities.

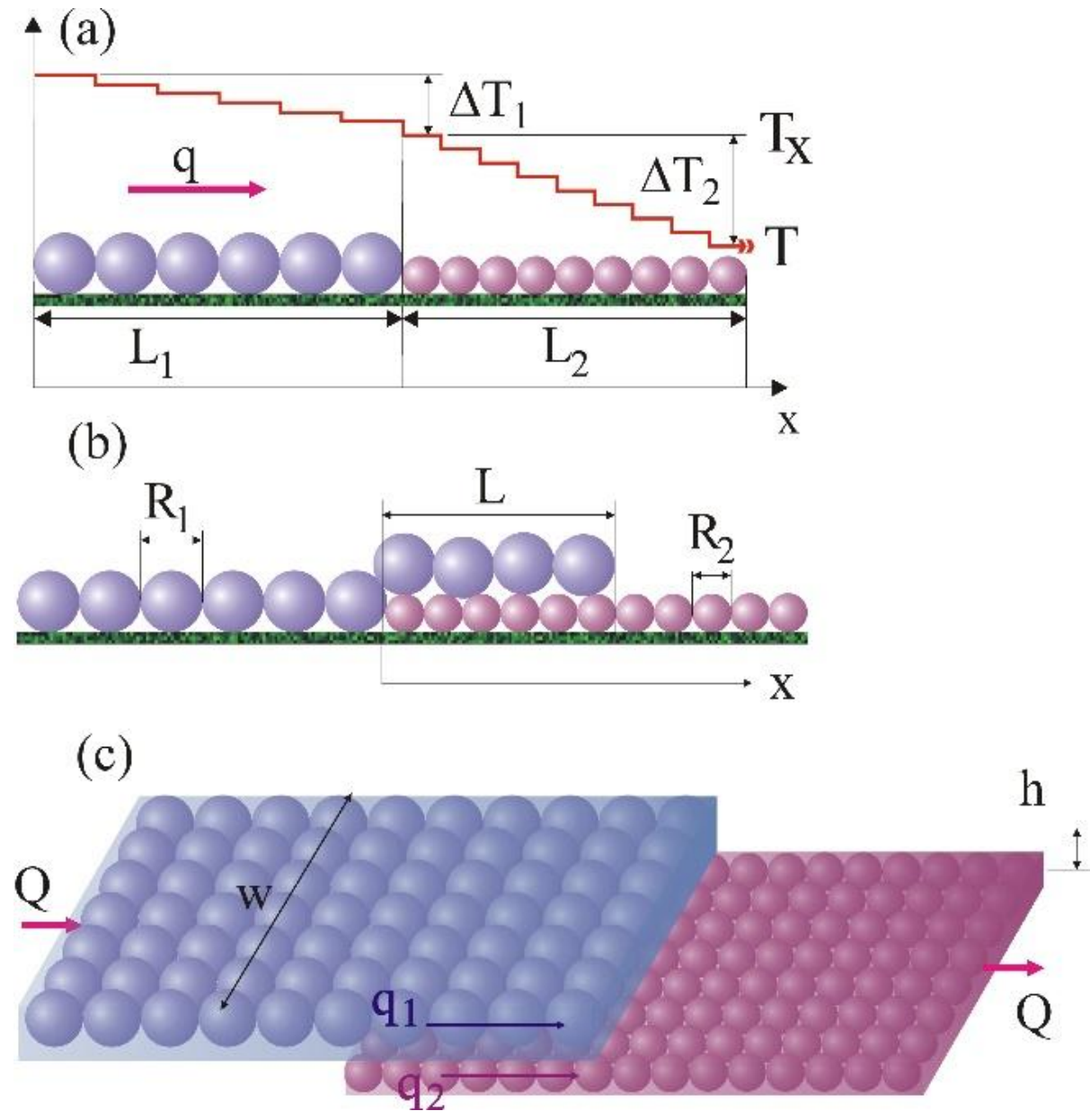

**Fig. 5.** *Different arrangements of nanospheres. (a) Two straight linear chains in series. Rot step line schematically shows temperature variation along the arrangement. (b and c) Midline cross section (b) of two overlapping layers from spheres with different radii. L is the length of the overlap region along the heat flow. Overlapping layers in three dimensions (c). Each layer represents a square array of identic nanospheres. Symbols are explained in the text.*

Consider two chains of lengths $L_1$ and $L_2$, composed of nanospheres of radii, $R_1$ and $R_2$, respectively, connected in series (Fig. 5a). All inter-sphere contact areas are identical in both chains. Under a temperature difference $\Delta T$ applied across this combined chain, the steady-state heat flux $q$ is the same for both chains, causing temperature differences $\Delta T_1$ and $\Delta T_2$, respectively. Neglecting contribution of the temperature step at the single contact between sub-chains ensures $\Delta T = \Delta T_1 + \Delta T_2$, so that the temperatures at the chain ends are $T$ and $T + \Delta T = T + \Delta T_1 + \Delta T_2$ (at the outer right and left ends, respectively). Here again $T_\mathrm{x} = T + \Delta T_2$ is an unknown temperature at the contact between sub-chains. We assume that the temperature difference between adjacent spheres in both sub-chains is small compared to temperatures of these spheres. Then the thermal conductivity of a chain equals the thermal conductivity of a dimer from two identical spheres (Fig. 4) as it is given by Eq. (2) with $R_1 = R_2$ and $\delta T \ll T$. Due to continuity, we can express the reverse net heat flux via parameters of each chain separately:

$$q_{1\to 2} = \frac{1}{L_2}\int_T^{T_\mathrm{x}} \mathrm{d}t\, \kappa_2(t) = \frac{1}{L_1}\int_{T_\mathrm{x}}^{T+\Delta T} dt\, \kappa_1(t). \tag{3}$$

Here $\kappa_1$ and $\kappa_2$ are thermal conductivities of chains composed from spheres with radii $R_1$ and $R_2$, respectively. Since $\Delta T$ is not necessarily small compared to $T$, integral Eq. (3) has to be solved numerically to find $T_\mathrm{x}$ and then $q_{1\to 2}$. The forward heat flux, $q_{2\to 1}$, is computed in a similar way albeit with a different $T_\mathrm{x}$.

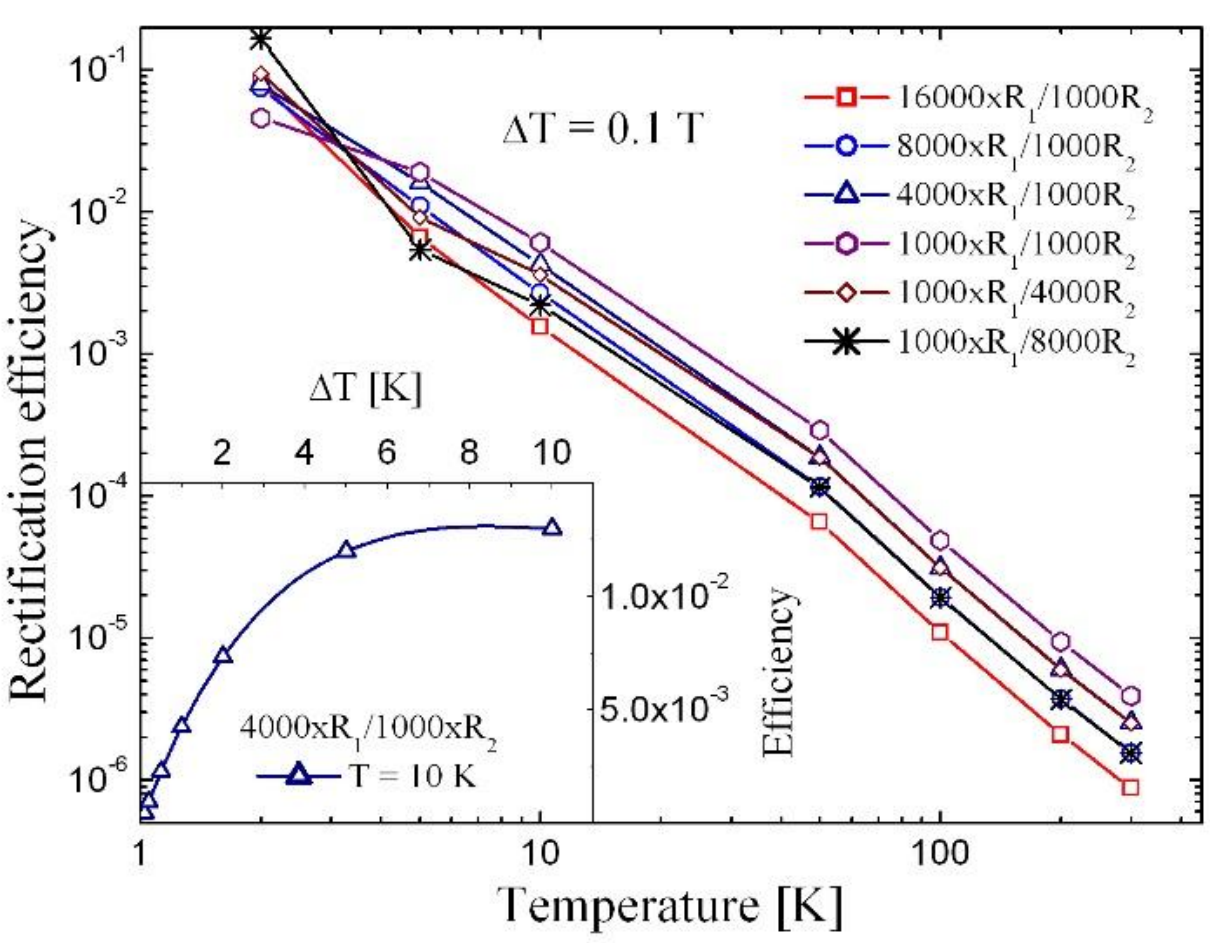

**Fig. 6.** *Rectification efficiencies (according to Eq. (2.1) for pairs of two sequential nanosphere sub-chains with different lengths. Radii of nanospheres in sub-chains are* $R_1 = 40$ *nm and* $R_2 = 10$ *nm. Sub-chain lengths are given in units of sphere radii. Temperature difference* $\Delta T$ *over the combined chain equals one tenth of the temperature at the lower end. Inset shows the effect of the increasing* $\Delta T$ *at T = 10 K for a chain from 4000* $R_1$ *spheres and 1000* $R_2$ *spheres.*

Rectification efficiencies for chains of nanospheres are presented in Fig. 6. Values are computed with Eq. (2.1) for sub-chains from spheres with radii $R_1 = 40$ nm and $R_2 = 10$ nm and different sub-chain lengths. Remarkably, efficiency increases at all temperatures with an increase in $\Delta T$ (Inset).

## 2. Rectification in overlapping nanosphere layers

Because the realization of linear chains of nanospheres is experimentally challenging, we consider a more realistic geometry of two partially overlapping layers composed of nanospheres of different radii. The geometry is depicted in Figs. 5b and 5c. Each layer is formed by identical nanospheres of different radii $R_1 > R_2$, arranged in a regular square array. Heat flows predominantly along each layer, while energy exchange between layers occurs through a distributed network of vertical contacts within the overlap region of length $L$. For simplicity, all inter-sphere contacts are assumed to have the same effective area S and transparency α. We consider linear regime $\delta T, \Delta T \ll T$ for all contacts and for the whole overlap and therefore neglect temperature dependence of thermal conductivities within the overlap.

With thermal conductivities defined in Eq. (2), the position dependent heat flux through a single layer of spheres with radius $R$ at temperature $T$ is given by Fourier's law $q(x) = -\kappa \frac{\mathrm{d}}{\mathrm{d}x} T(x)$ [W m$^{-2}$], while the total heat flow is $q(x) S \frac{w}{2R}$ . Here $S \frac{w}{2R}$ is the total cross-sectional area available for the horizontal heat flow within the layer, accounting for the single-contact area S and the number of contacts $\frac{w}{2R}$ across the layer of width $w$.

In the overlap region, heat fluxes in two layers, $q_1$ and $q_2$,change due to the heat "leak" between the layers. Within a length $\mathrm{d}x$, the "leak" heat flow from the layer of spheres with radius $R_1$ to the layer of spheres with radius $R_2$ can be written as $\mathrm{d}Q_{1\to2} = -\kappa_{1\to2}\ \delta T(x)/h\ S\, w\, \mathrm{d}x/(4\bar{R}^2)$ [W]. Here, the direction-dependent thermal conductivity $\kappa_{1\to2}$ (or $\kappa_{2\to1}$) accounts for rectification; $\delta T(x) = T_2 - T_1$ is the temperature difference between layer midplanes separated by $h \approx R_1 + R_2$. With the mean radius $\bar{R} = (R_1 + R_2)/2$, the factor $S\, w\, \mathrm{d}x/(4\bar{R}^2)$ represents the total cross-section available for the vertical heat flow within the overlapping region of length $\mathrm{d}x$ and width $w$.

For the flow direction shown in Fig. 5c, heat flows in the layers, $Q_1$ and $Q_2$, and the "leak" $\mathrm{d}Q_{1\to2}$ inside the overlapping region and the total heat flow, $Q$, outside of this region are connected by the obvious continuity relations: $\mathrm{d}Q_{1\to2} = -\mathrm{d}Q_1 = \mathrm{d}Q_2$ or $2\,\mathrm{d}Q_{1\to2} = \mathrm{d}Q_2 - \mathrm{d}Q_1$ (i) and $Q = Q_1 + Q_2 = q_1(x)S\frac{w}{2R_1} - q_2(x)S\frac{w}{2R_2}$ (ii). Heat transport in the overlapping region is then described by the following equation system

$$\frac{\mathrm{d}}{\mathrm{d}x}\delta T(x) = \frac{q_1(x)}{\kappa_1(T)} - \frac{q_2(x)}{\kappa_2(T)}$$

$$\frac{d}{dx}\left(\frac{q_1(x)}{R_1} - \frac{q_2(x)}{R_2}\right) = \frac{\kappa_{1\leftrightarrow2}}{4\bar{R}^3}\delta T(x)$$

$$\frac{q_1(x)}{R_1} + \frac{q_2(x)}{R_2} = 2\frac{Q}{w\,S}. \tag{6}$$

The first equation is obtained by differentiating $\delta T(x) = T_2 - T_1$, while the second and the third are continuity conditions (i) and (ii), respectively. Note that the contact area cancels out. Here $\kappa_{1\leftrightarrow2}$ is the direction-dependent vertical thermal conductivity presented by Eq. (3). One recognizes that this system is mathematically analogous to the Telegraph equations [22] for an overdamped transmission line at zero frequency (DC) regime after the substitutions $q$ for $j$ (current density) and $\delta T$ for $U$ (voltage).

With the boundary conditions $q_1(0) = \frac{2QR_1}{S\,w}$; $q_1(L) = 0$ and $q_2(L) = \frac{2QR_2}{S\,w}$; $q_2(0) = 0$, the solutions for heat fluxes become

$$q_1(x) = 2\kappa_1\frac{Q_0}{w\,S}\frac{R_2R_1}{\kappa_1R_2+\kappa_2R_1}\left(1 + \frac{\kappa_2R_1}{\kappa_1R_2}\frac{\sinh[\gamma(L-x)]}{\sinh(\gamma L)} - \frac{\sinh(\gamma x)}{\sinh(\gamma L)}\right),$$

$$q_2(x) = 2\kappa_2\frac{Q_0}{w\,S}\frac{R_2R_1}{\kappa_1R_2+\kappa_2R_1}\left(1 + \frac{\kappa_1R_2}{\kappa_2R_1}\frac{\sinh(\gamma x)}{\sinh(\gamma L)} - \frac{\sinh[\gamma(L-x)]}{\sinh(\gamma L)}\right), \tag{7}$$

where the damping constant $\gamma = \left[\kappa_{1\leftrightarrow2}\frac{(\kappa_1R_2+\kappa_2R_1)}{4\bar{R}^3\kappa_1\kappa_2}\right]^{\frac{1}{2}}$ has dimensions of inverse length. The averaged thermal conductivity of the overlapping region to the first order is

$$\bar{\kappa} = \frac{\kappa_1R_2+\kappa_2R_1}{\bar{R}}\frac{2}{1+\frac{1}{\gamma\,L\sinh(\gamma L)}\left[\left(\frac{\kappa_1R_2}{\kappa_2R_1}+\frac{\kappa_2R_1}{\kappa_1R_2}\right)\cosh(\gamma L)+2\right]}. \tag{8}$$

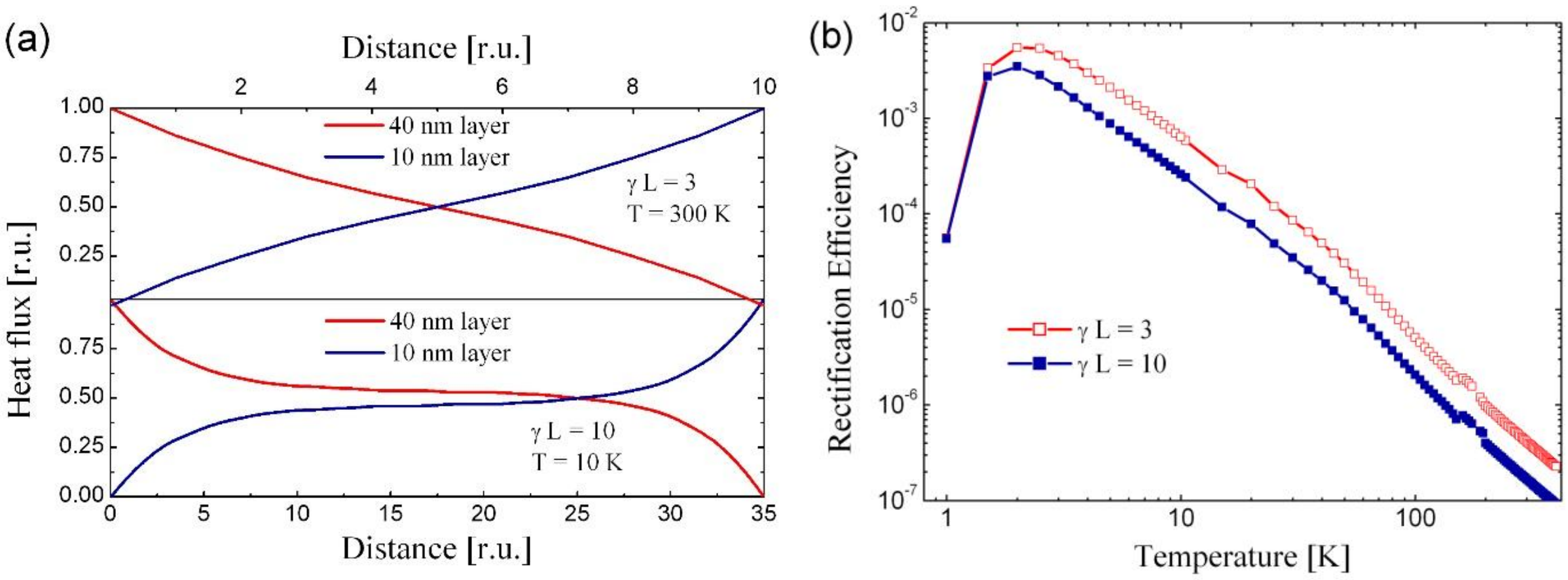

**Fig. 7.** *Heat fluxes in overlapping layers as function of position along the heat flow (a) and rectification efficiencies (b) for overlapping layers from Ge nanospheres with radii 40 and 10 nm. The dimensionless lengths of the overlapping region are $\gamma L \approx 3$ and $\gamma L \approx 10$. Data are obtained with Eqs. (2.1), (7) and (8).*

Fig. 7a shows heat fluxes through the upper ($R_1$ = 40 nm) and bottom ($R_1$ = 10 nm) layers for two $\gamma L$ values. The exchange of fluxes between layers occurs within a few spheres. For large $\gamma L$, heat fluxes in the middle of the overlapping region remain constant. Hence, this region is excluded from rectification. Fig 7b depicts rectification efficiencies of the overlapping region computed with Eq. (2.1) and (8) for the same two $\gamma L$ values.

The expression for $\bar{\kappa}$ is symmetric with respect to an interchange of $\kappa_1, R_1$ and $\kappa_2 R_2$ as it is required by the symmetry of the problem in the absence of rectification. Rectification is entirely controlled by $\kappa_{1\leftrightarrow 2}$ in the damping constant $\gamma$. With increasing overlap length $L$, the rectification efficiency decreases and approaches direction independent value $\frac{\kappa_1 R_2 + \kappa_2 R_1}{R_2 + R_1}$. For $\gamma L \ll 1$, as expected, $\bar{\kappa} \to \kappa_{1\leftrightarrow 2}$. In this case the thermal conductivity of the overlap reduces to that of a single contact between different nanospheres.

## CONCLUSION

We have demonstrated that thermal rectification in nanostructured materials can arise purely from phonon confinement in nanoparticles of identical material but different size. In the proposed mechanism, geometrical confinement suppresses long-wavelength phonons more effectively in smaller particles, creating a direction-dependent spectrum of phonon flux through the contact between unequal nanospheres. This asymmetry leads to different heat fluxes for equal but opposite temperature gradients and hence to thermal rectification without requiring material heterogeneities (e.g. chemical heterogeneity) or strong nonlinearities.

Using an analytical model in the Casimir regime, we derived expressions for heat transport and rectification efficiency for three experimentally relevant geometries: a single contact between two nanospheres, sequential chains, and partially overlapping nanoparticle layers. We identified two rectification mechanisms: direct asymmetry of the confined phonon spectra at a single contact, and size-dependent nonlinear temperature behavior of effective thermal conductivities in extended systems. The model predicts rectification efficiencies of a few tens of percent for a single contact between two different nanospheres and for nanosphere chains with radii of a few tens of nanometers.

These results show that size-induced phonon confinement provides a simple and scalable mechanism for directional heat transport in nanoparticle-based materials and granular nanostructures.

**Conflict of Interest**

ADS, MS, AZ and MDV filed a patent based on this work.

**Author contributions**

ADS developed the physical model and mathematical approach, ADS and MS performed computations, ADS, MS, AZ and MDV analyzed computational results, MDV expired the work via communicating experimental results prior to publication, ADS wrote the manuscript with contributions from all authors.

**Appendix A: Material Parameters used for computations**

All numerical results presented in the main text are computed for nanospheres composed of crystalline germanium. Material parameters used are summarized in Table A1 and typical for various dielectric crystals listed in Table A2

The Debye temperature $T_D$ is matched to its experimental value by choosing $a_0 = 0.33$ nm, corresponding to the crystalline structure of Ge with eight atoms in each unit cell. The experimental heat capacity corresponds to the Deby model up to room temperature. The Debye model along with the Kinetic equation approach reasonably well describe thermal conductivity of Ge up to temperatures of a few tens of K. At higher temperatures, anharmonicity (phonon-phonon scattering) reduces phonon mean free path and that diminishes the thermal conductivity. However, $l_{MPF}$ remains in the micrometer range even at room temperatures ensuring the Casimir regime of phonon propagation in nanospheres. At 400 K, the mean phonon wavelength of 0.6 nm is comparable to the size of the unit cell that causes saturation in the temperature dependence of thermal conductivity in chain of nanospheres.

Table A1. Material parameters of pure Ge used in the model. $T_D$ is the experimental Debye temperature [23]. $c_V$ and $\kappa$ are experimental heat capacity and thermal conductivity, respectively (at 6 and 2 K, respectively). Values computed in the framework of the Debye model are indexed with "D".

| $u_l$ | $u_t$ | $a_0$ | $T_D$ | $c_V$ | $\kappa$ | $c_D$ | $\kappa_D$ |
|---|---|---|---|---|---|---|---|
| m sec$^{-1}$ | m sec$^{-1}$ | nm | K | J m$^{-3}$ K$^{-1}$ | W m$^{-1}$ K$^{-1}$ | J m$^{-3}$ K$^{-1}$ | W m$^{-1}$ K$^{-1}$ |
| $5.3\times10^3$ | $3.45\times10^3$ | 0.57 | 374 | 495 | 70 | 480 | 69 |

Within our model, thermal rectification is determined primarily by geometric confinement rather than material-specific properties. Table A2 compares rectification efficiency (computed with Eq. 2.1) for several dielectric materials: at 10 K for a dimer with $R_1$ = 40 nm and $R_2$ = 10 nm, demonstrating quantitatively similar values to those for Ge (Fig. 4).

Table A2. Material parameters and rectification efficiency at $\delta T = T = 10$ K for nanosphere dimers with $R_1$ = 40 nm and $R_2$ = 10 nm.

| Material | $u_l$<br>m s$^{-1}$ | $T_D$<br>K | Rectification efficiency<br>Rec (10 K) | Refs |
|---|---|---|---|---|
| Ge | $5.3\times10^3$ | 374 | 0.017 | [23] |
| Si | $9.1\times10^3$ | 645 | 0.042 | [21] |
| $TiO_2$ | $8.8\times10^3$ | 560 | 0.04 | [24,25] |
| $HfO_2$ | $5.9\times10^3$ | 480 | 0.02 | [26] |